# Unveiling the landscape of Mottness and its proximity to superconductivity in 4Hb-TaS$_2$

Ping Wu[1,2,#], Zhuying Wang[1,#], Yunmei Zhang[1,#], Ziyan Chen[3,#], Shuikang Yu[1,4], Wanru Ma[1], Min Shan[5], Zeyu Liang[1], Xiaoyu Wei[1], Junzhe Wang[1], Wanlin Cheng[1], Zuowei Liang[5], Xuechen Zhang[1,4], Tao Wu[1,4,5,6], Yoshinari Okada[2], Kun Jiang[3,*], Zhenyu Wang[1,4*] and Xianhui Chen[1,4,5,6*]

[1]*Department of Physics, CAS Key Laboratory of Strongly-coupled Quantum Matter Physics, University of Science and Technology of China, Hefei, Anhui 230026, China*
[2]*Quantum Materials Science Unit, Okinawa Institute of Science and Technology (OIST), Okinawa 904-0495, Japan*
[3]*Beijing National Laboratory for Condensed Matter Physics and Institute of Physics, Chinese Academy of Sciences, Beijing 100190, China*
[4]*Hefei National Laboratory, University of Science and Technology of China, Hefei 230088, China*
[5]*Hefei National Research Center for Physical Sciences at the Microscale, University of Science and Technology of China, Hefei, 230026, China*
[6]*Collaborative Innovation Center of Advanced Microstructures, Nanjing 210093, China*

[#]These authors contributed equally to this work
[*]Correspondence and requests for materials should be addressed to Z.W. (zywang2@ustc.edu.cn), K. J. (jiangkun@iphy.ac.cn) or X.-H.C. (chenxh@ustc.edu.cn).

**Mott physics is at the root of a plethora of many-body quantum phenomena in quantum materials. Recently, the stacked or twisted structures of van der Waals (vdW) materials have emerged as a unique platform for realizing exotic correlated states in the vicinity of the Mott transition. However, the definitive feature of Mottness and how it rules the low-energy electronic state remain elusive and experimentally inaccessible in many interesting regimes. Here, we quantitatively describe a filling-controlled Mott state and its interplay with superconductivity by scanning tunnelling spectroscopy in a vdW bulk heterostructure, 4Hb-TaS$_2$, that interleaves strongly correlated 1T-TaS$_2$ layers with superconducting 1H-TaS$_2$ layers. The fine tunability of electron doping induced by interlayer charge transfer allows us to continuously track the spectral function with unsurpassed energy resolution from a depleted narrow band (0.2 electrons per site) toward a Mott transition at half filling. The gradually emerging Mott-Hubbard bands, followed by the sharpening and vanishing of the central quasiparticle peak as predicted in the Brinkman-Rice scenario, unambiguously demonstrate the Mott physics at play. Importantly, the renormalization of the low-energy electrons acts destructively on the superconducting pairing potential, leaving behind nonsuperconducting, paramagnetic puddles at the nanoscale. Our results reveal a seminal system near the border of the Mott criterion that enables us to illustrate the predictive power of the Hubbard model, and set such heterostructures as promising ground for realizing new correlated states in the heavily doped Mott regime.**

Mott insulator is a paradigmatic example to study the relevant role of electron correlations. It represents the localization of electrons at half band-filling when the interaction energy of electrons--characterized by the Coulomb repulsion $U$-- dominates over their kinetic energy characterized by the bandwidth $W$ [1,2]. The vicinity of such Mott transitions is believed to host exotic many-body quantum states, including unconventional superconductivity [3,4], quantum spin liquids [5,6] and strange metallicity [7]. Although many efforts have been made to understand the microscopic basis of the Mott physics, key features in many interesting doping regimes remain difficult to access experimentally.

Recently, two-dimensional van der Waals (vdW) materials have emerged as highly tunable, strongly correlated electronic systems by twisting or hetero-stacking, thus opening a new avenue for exploring the landscape of Mott physics and realizing correlated states that are not foreseeable from single layers alone [8-16]. The 4Hb polytype of tantalum disulfide (4Hb-$TaS_2$) is one case in point [17]: it is a naturally occurring bulk vdW heterostructure composed of alternating 1T- and 1H-$TaS_2$ layers (Fig. 1a). The 1H layer is superconducting and features spin-valley locking of the Ising type [18,19]. The 1T layers host a strong $\sqrt{13} \times \sqrt{13}$ superstructure under which every 13 Ta atoms are reconstructed into a Star of David (SoD) cluster, each holding one Ta-$5d$ electron localized on the central Ta ion. These localized electrons are believed to bear large onsite Coulomb interactions compared with their intralayer hopping amplitude, therefore giving rise to a Mott insulator and a candidate quantum spin liquid on a triangular lattice [20-22]. However, such Mott state is susceptible to out-of-plane stacking in bulk 1T-$TaS_2$, where the interlayer dimerization of the SoD clusters leads to a band insulator with an even number of electrons in the Ta-$5d$ orbitals [23-26].

In 4Hb-$TaS_2$, the interleaving of 1H superconducting layers can eliminate interlayer dimerization but at the expense of suffering marked electron transfer from the 1T layer to the 1H layer. This results in an almost empty narrow band in the 1T layer [27,28], and consequently, there is little room for Mottness or other strongly correlated effects. Nevertheless, the superconductivity in 4Hb-$TaS_2$ exhibits a series of highly unconventional features, including signatures of broken time-reversal symmetry [29] and multicomponent order parameters [30,31], topological edge modes [32], residual specific heat [29], and a puzzling magnetic memory effect manifested in the spontaneous vortex phase [33]. Most of these phenomena occur in lightly Se-doped samples and imply the critical role of electron correlations. In comparison, the spectroscopy data for non-superconducting 1T/1H-$TaS_2$ bilayer films exhibit large a Mott gap feature together with a narrow zero-bias peak, which has been interpreted in terms of local moment in the 1T layer being Kondo-screened by the itinerant electrons of the 1H layer [18,34-36]. A recent first-principles study puts forward an alternative explanation that diminishes the role of screening by the 1H electrons and suggests a doped Mott insulator whose doping level depends on the distance between the layers [37]. This raises two central questions in 4Hb-$TaS_2$: first, how does the correlation effect develop for the 1T-electrons, specifically, to what extent does Mott physics govern the low-energy states, and second, how do such correlations subsequently influence superconductivity? The absence of a clear characterization of the electronic state has led to a strong dependence on theoretical input to clarify the underlying physics, which calls for systematic spectroscopy studies.

Here, we use scanning tunneling microscopy and spectroscopy (STM/S) to reveal the fate of 1T-electrons in 4Hb-TaS$_2$. Taking advantage of the subtle charge transfer between the 1T and 1H layers, we successfully track the evolution of the single-particle spectral function continuously from 0.2 electrons per SoD cluster toward half filling with an unsurpassed energy resolution. Systematic measurements and analyses enable us to quantitatively determine the emerging Mott-Hubbard gap along with the diminishment of the central quasiparticle peak. These observations, taken together with our dynamical mean-field theory (DMFT) calculation, place the system near the border line of the Mott criterion. Moreover, we provide spectroscopic evidence that this renormalization of the quasiparticle states locally alters the superconducting gap potential and gives rise to nonsuperconducting puddles with local moment in the 1T layer.

Single crystals were synthesized with a low selenium concentration (4Hb-TaS$_{1.99}$Se$_{0.01}$) as described in previous reports [18, 27-33,]. The transport characterization is shown in Supplementary Fig.1, which confirms the $\sqrt{13} \times \sqrt{13}$ charge density wave (CDW) transition at 315 K and the superconducting transition at 3.0 K. Cleaving naturally exposes the sulfur surfaces of both the 1T and 1H terminations, and these two types of surfaces can be distinguished morphologically and spectroscopically. For the 1H termination, both the atomic and 3 × 3 CDW periodicities are apparent in the STM topography (Fig.1b). The differential conductance (*dI/dV*) spectrum shows a typical V-shaped metallic character (Supplementary Fig.2) and the opening of a major superconducting gap $\Delta_{SC}^{1H} \approx 0.5\ meV$ at low temperature (inset of Fig.1b). Notably, the strong $\sqrt{13} \times \sqrt{13}$ superlattice characteristic of the 1T layer (Fig. 1c) is also imprinted onto the 1H layer, indicating a remarkable interaction between adjacent layers. For the 1T termination, the majority of the *dI/dV* spectra exhibit an electron–hole asymmetric peak with its center slightly above the Fermi energy (E$_F$), as shown in Fig. 1d. This peak corresponds to the isolated narrow band of the Ta-5*d* electrons with the $\sqrt{13} \times \sqrt{13}$ CDW order, as supported by our first principal calculation (Fig. 1e), which has shifted above E$_F$ owing to charge transfer [27, 28]. The peak width at half height allows us to estimate the bandwidth of this narrow band, where we find $W_0 \approx 80\ meV$. By integrating the spectral weight below E$_F$ and calculating its proportion to the total weight of this narrow band, we find that approximately 0.2 electrons remain in each SoD cluster. These observations set a good starting point for further exploration.

In contrast to the homogeneous metallic state of the 1H layer, the low-energy spectral feature exhibits high diversity among different SoD clusters in the 1T layer. For example, Fig. 2a shows an STM topography with a 100-by-100-nm field of view (FOV), which contains approximately four thousand SoD clusters. The obtained electronic spectrum at the center of each cluster reveals that, in addition to the commonly observed depleted narrowband feature, approximately 12% of the clusters show a prominent zero-bias peak flanked by two broader peaks, reminiscent of the Kondo resonance observed in 1T/1H-TaS$_2$ bilayer films [16,34-36] and consistent with a previous report [28]; a series of intermediate states are also observed, as shown in Fig. 2b. Through *dI/dV* mapping, we found that these low-energy states are distributed in the *center* of the SoD clusters, whereas between the clusters, they exhibit depleted narrowband features despite weak intensity (Supplementary Fig.3). This in-phase distribution suggests that these varying low-energy electronic features should stem from the narrow band, whose electron filling somehow changes among different clusters. In fact, their electron-doping nature can be directly revealed by the spectral line shapes at relatively high energies. For example, focusing on the broad peak located at approximately 0.5 eV below $E_F$, as shown in Fig.2c, we find that it progressively shifts to negative binding energies, from clusters showing the depleted narrow band to those with zero-bias peaks. As such, this direct link clarifies that the low-energy spectral function reflects an evolution of the narrow band upon increasing electron filling.

Next, we turn to the origin of such electron doping. In these 4Hb-TaS$_2$ samples, 0.5% of the sulfur atoms were replaced by selenium to stabilize the crystal. If we assume that each SoD cluster (13 Ta and 26 sulfur atoms) contains no more than one selenium substitution (which is reasonable due to the ultralow Se concentration), approximately 13% of the clusters are influenced. Se defects can inhibit charge transfer to the host clusters, as supported by a recent calculation [38]. This hypothesis is further confirmed by two controlled experiments (Supplementary Note 4): the proportion of clusters hosting a zero-bias peak significantly increases in samples with higher Se concentrations, whereas it decreases when uniaxial tensile strain is applied to the sample, which in principle reduces interlayer separation and promotes charge transfer. Therefore, the electron filling level in each cluster is finely controlled through the charge transfer at a subtle balance of selenium substitution and interlayer spacing. This saturation is unique and allows an in-depth study of the filling-controlled electronic state here. More importantly, the doping effect deviates from that of conventional metallic systems and shows strong spatial inhomogeneity, as in cuprates [39]. This feature comes from the weak screening of electron Coulomb potential in a strongly correlated system.

In Fig. 2d, we have sorted more than one hundred typical *dI/dV* spectra according to the exact energy of the broad hump near -0.5 eV. We note that the involvement of these high-energy spectral features is essential for the analysis: first, they are reliable indicators of the actual amount of electron doping, and second, their spectral weights offer a basis for normalizing the *dI/dV* spectra, thus enabling a meaningful comparison of the low-energy density of states (DOS) among different clusters. The resulting false color plot (Fig. 2d) depicts a full picture of how the nearly empty narrow band, with increasing electron filling (note the collective shift of the high-energy band features), transforms into two shallow sub-bands below and above $E_F$ along with the sharpening and vanishing of the central peak. For better visualization, several representative spectra are plotted in Fig. 3a, which capture the main characteristics of the evolution. The symmetrical transfer of spectral weight from the original coherent part (the narrow Ta-5*d* band) to the incoherent part (upper and lower Hubbard bands; abbreviated as UHB and LHB), at first glance, follows the prediction for a Mott transition, which will be tested in great detail later.

The unprecedented full set of spectroscopic data with high energy resolution enables us to extract key parameters for the theoretical models with which the underlying physics can be revealed. To proceed, we focus our analyses on the energy range of $\pm 0.25$ eV, which includes the most relevant information of the isolated narrow band. The total spectral weight in this range, obtained by integrating the local DOS, remains roughly unchanged with electron filling (Fig. 2e). However, the narrowband feature becomes sharper upon doping, with its maximum pinned near $E_F$; its intensity then decreases when the UHB and LHB start to develop (Fig. 3a). In this region, we can extract several key parameters, including the onsite Coulomb *U* and positions of the LHB ($\varepsilon$) and UHB ($\varepsilon + U$), as well as the widths and spectral weights of the Hubbard bands and the central peak, by fitting these three features with Lorentzian line shape (Fig.3b and Supplementary Fig.5). Now, we plot the results versus electron filling; this is done by employing $\varepsilon$ as a measure of electron doping, since we find that it has a linear relationship with the position of high-energy features (Supplementary Fig.6). The first direct observation is that while both LHB and UHB shift in energy with doping, a constant onsite $U \cong 190 \pm 10 meV$ is found in the entire dataset (Fig.3c). A similar value was found for the AC stacked termination in bulk 1T-TaS$_2$ (Supplementary Fig.7). When the LHB overlaps with the central peak, it carries more spectral weight than the UHB does, but the difference decreases with electron doping and almost disappears toward half-filling (Fig. 3d). Concurrently, the central peak becomes sharper (as indicated by its full width at half maximum, FWHM; Fig. 3f) and vanishes (its spectral weight, Fig. 3g), as it undergoes significant renormalization by onsite $U$. The overall evolution of the spectral function can be well captured by our DMFT calculations with quantum Monte-Carlo (Fig. 3g) and numerical renormalization group solvers (Supplementary Fig. 8). Although our simplified model calculation deviates from the tunneling peaks due to the solver and approximations, it is fruitful to compare the theoretical results with our experimental data to look for the general signatures of doped Mott systems.

With the knowledge of ($\varepsilon, U, W_0$), we are ready to discuss the nature of the central peak observed here, i.e., whether it is a Kondo resonance with the screening of 1H-electrons or the remanent coherent state of 1T-electrons in the Mott picture. For a single-impurity Kondo problem, the Kondo temperature can be calculated with the Anderson impurity model as $k_B T_k = \omega W_0 \sqrt{U/4W_0} \exp(-\pi |\varepsilon||\varepsilon + U|/W_0 U)$ with $\omega = 0.4128$ being the Wilson number. Since all the microscopic parameters are experimentally accessible, $T_k$ can be directly calculated and is found to be approximately 40 K in the case of near-half filling. Here, the breakdown of the Kondo explanation is evident from the width of the central peak at ultralow temperatures. In fact, the half width at half maximum (HWHM) of a Kondo resonance follows a famous expression of $\Gamma = \frac{1}{2}\sqrt{(\alpha k_b T)^2 + (2 k_b T_k)^2}$ and is irrelevant to $T$ when $T \ll T_k$ [40-42], which reflects the finite temperature scale of a Kondo state. However, we find that the HWHM of the central peak observed here is generally an order of magnitude smaller, does not saturate but decreases linearly with temperature from 4.2 to 0.3 K (Fig.3h), which suggests an essential mechanism at zero temperature (Supplementary Fig.9). The significantly reduced energy scale of this peak, together with the fact that it always connects to the LHB, leads us to attribute it to the remnant, narrowing coherent band as in the Brinkman–Rice picture of Mott transitions [43], where the vanishing width of the central peak indicates the divergence of the effective mass of electrons approaching half-filling. Therefore, our data demonstrate a doped Mott state [44-46] in the 1T layer of 4Hb-TaS$_2$, and the 1H layers primarily act as a charge reservoir [37]; the ratio of $U/W_0$ ~2 places the system near the border line of the Mott criterion.

The bulk superconductivity of 4Hb-TaS$_2$, in which many fascinating phenomena have been reported [29-33], further provides a unique opportunity to study the manifestation of a doped Mott state in proximity to superconducting layers. To establish a benchmark, we first examine the superconductivity of the 1H layer. The measured $dI/dV$ spectrum shows a homogeneous superconducting gap with a U-shaped bottom (Fig. 4a), and no visible Yu-Shiba-Rusinov states are found. When moving to the 1T layer, we find a relatively weak, electron-hole symmetric gap near $E_F$. This smaller gap exhibits a V-shaped bottom and disappears at temperatures above Tc and at out-of-plane magnetic fields of 2 T (Fig.4b), clarifying its identity as a superconducting gap possibly induced by interlayer proximity. This superconducting gap on the 1T layer has not been reported before. Interestingly, we find that the gap size $\Delta_{SC}^{1T}$ dramatically varies between SoD clusters (Fig.4d), being largely constant inside each of them. It takes a maximum value of $\Delta_{SC}^{1T} \approx$ 0.35 meV on clusters featuring a depleted narrow band and collapses to zero ($\Delta_{SC}^{1T} \approx 0$) on those showing Hubbard bands. To better visualize the overall trends, we present the tens of low-energy spectra of different clusters in Fig. 4e after sorting them by the electron doping level (see wide-energy spectra in Supplementary Fig.9). One can see that with increasing electron doping, the superconducting gap gradually decreases and finally vanishes when LHB and UHB develop, leaving behind only the sharp metallic peak at $E_F$. As discussed in the Brinkman-Rice scenario, the hopping of low-energy quasiparticles is strongly renormalized owing to the large energy cost of changing electron occupation. The coupling between 1T and 1H also strongly depends on this renormalization. Therefore, if the superconductivity in the 1T layer comes from the proximity effect, the superconducting gap becomes weaker when the correlation effect of the SoD clusters becomes stronger, which is consistent with our experimental findings.

The discovery of correlation-dressed superconductivity has multiple ramifications. First, it results in a spatially inhomogeneous superconducting state in the 1T layer, and approximately 12% of the SoD clusters therein do not show a superconducting feature but instead exhibit a sharp zero-bias peak. Second, we characterized the central quasiparticle peak under external magnetic fields and found that it splits into two peaks with increasing magnetic field (Supplementary Fig. 10). The linear field dependence can be attributed to Zeeman splitting with a Landé factor of g ~2 for the localized electrons in these clusters (with small bandwidths after renormalization). Therefore, these nonsuperconducting clusters can be regarded as a unique type of paramagnetic "impurity" formed by electron correlations. These paramagnetic clusters could give rise to a negative Josephson coupling between the neighboring 1H and 1H' layers (which are rotated 180 degrees relative to each other; Fig. 1a) when the spin-valley locking is considered [47]. This provides a possible explanation for the unusual π phase shifts in the Little-Parks experiments [30]. Third, the SoD clusters hosting local moments seem to exhibit spatial correlations, as they are locally arranged with a short-range order in real space (Supplementary Fig.11). This suggests that although the proportion of these clusters is determined by Se substitution, their precise locations might be slightly redistributed by the nonlocal Coulomb interaction [48,49]. It remains to be seen whether any unconventional magnetic interactions can be established on the basis of such a spatial pattern that might be essential for the puzzling magnetic memory effect in 4Hb-TaS$_2$ [33].

In closing, we have made use of subtle interlayer charge transfer to establish an ensemble of doped Mott insulators with a large variety of electron fillings in the 1T layer of 4Hb-TaS$_2$. This allows us to track the evolution of the single-particle spectral function continuously from 0.2 electrons per SoD cluster toward half filling. Our spectroscopic data with unsurpassed energy resolution provide the first full characteristics of how a narrow band transforms into a Mott state by filling control, including the emergence of Mott-Hubbard bands and the narrowing of coherent peaks as in the famous Brinkman-Rice scenario. Remarkably, such renormalization ramifies the superconducting state of the 1T layer via locally closing the pairing gap and creating paramagnetic clusters, which sets the ground for understanding the unusual phenomena of the superconducting phase. Therefore, such vdW heterostructures provide new grounds to enrich the paradigm of doped Mott physics and represent exciting perspectives for engineering exotic quantum states.

# Figure 1

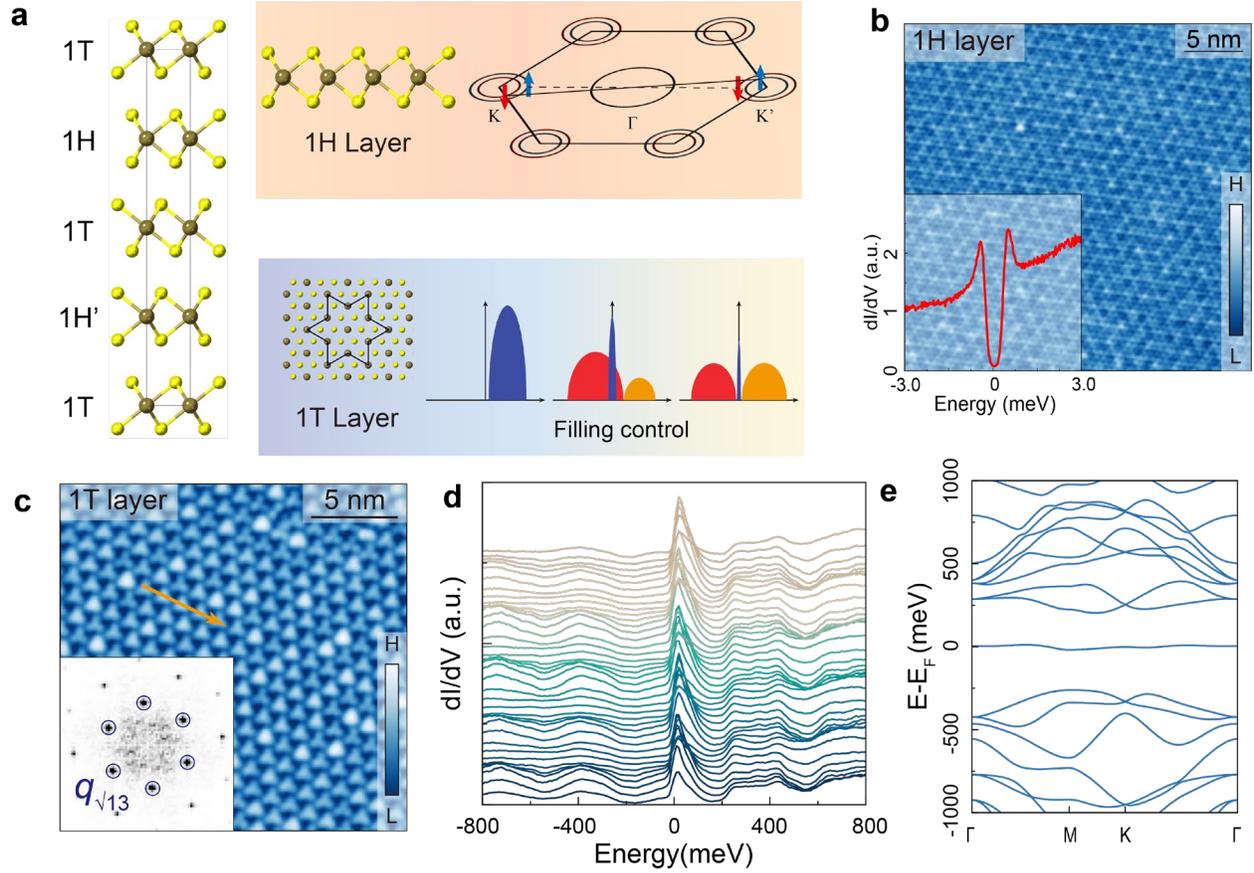

**Fig. 1 Atomic structure, charge density wave and depleted narrow band in 4Hb-TaS$_2$. a**, Crystalline structure of 4Hb-TaS$_2$, which interleaves strongly correlated 1T-TaS$_2$ layers with superconducting 1H-TaS$_2$ layers. **b**, Topography of the 1H termination showing 3 × 3 CDW modulation. Insert: superconducting gap obtained at 0.3 K. **c**, Topography of the 1T layer showing $\sqrt{13} \times \sqrt{13}$ CDW where every 13 Ta atoms are reconstructed into a SoD cluster. The inset shows its Fourier transform, in which the CDW peaks are marked by circles. **d**, dI/dV spectra obtained along the line shown in c. **e**, Calculated band structure of monolayer 1T-TaS$_2$ with $\sqrt{13} \times \sqrt{13}$ CDW. An isolated narrow band can be found at E$_F$. STM setup conditions: **b**: $V_s$ = -10 meV, $I_t$ = 1 nA. **c**: $V_s$ = -500 meV, $I_t$ = 1 nA. **d:** $V_s$ = -800 mV, $I_t$ = 2 nA, $V_{mod}$ =10 meV.

# Figure 2

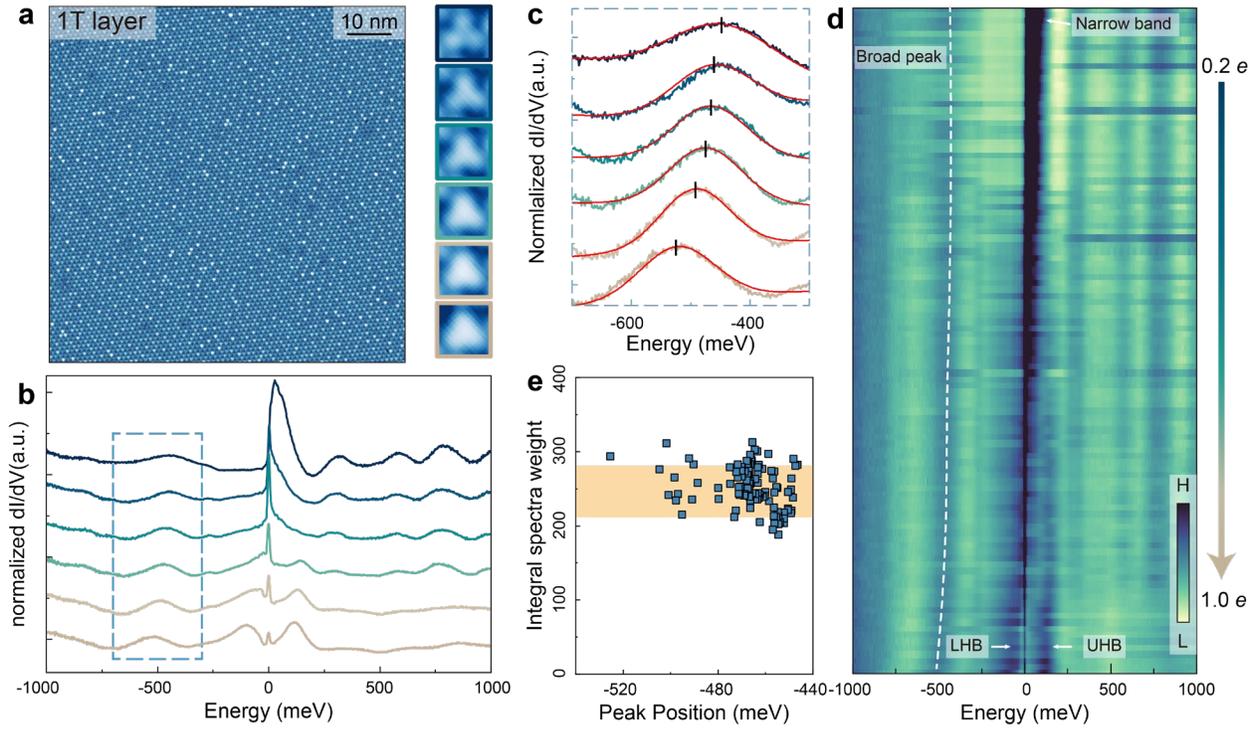

**Fig. 2 Evolution of the narrow band with varying electron filling. a**, Topographic image of a large FOV at the 1T termination. The apparent shapes of the SoD clusters exhibit high diversity owing to their different spectral features. **b**, Representative dI/dV spectra obtained at the centers of different SoD clusters. **c**, Energy shift of the broad peak (at approximately 0.5 eV below $E_F$) for the spectra in b, indicating an increase in the electron filling of each cluster from top to bottom. **d**, A false colormap of the normalized dI/dV spectra obtained at the center of one hundred clusters, which are sorted in according to the exact energy of the broad hump near -0.5 eV. **e**, Integral of the density of states from -250 meV to 250 meV in d. STM setup conditions: **a**: $V_s$ = -500 meV, $I_t$ = 1 nA; **b, c**: $V_s$ = -1000 mV, $I_t$ = 2.5 nA, $V_{mod}$ = 2; **d**, $V_s$ = -1000 mV, $I_t$ = 2.5 nA, $V_{mod}$ = 2 meV.

# Figure 3

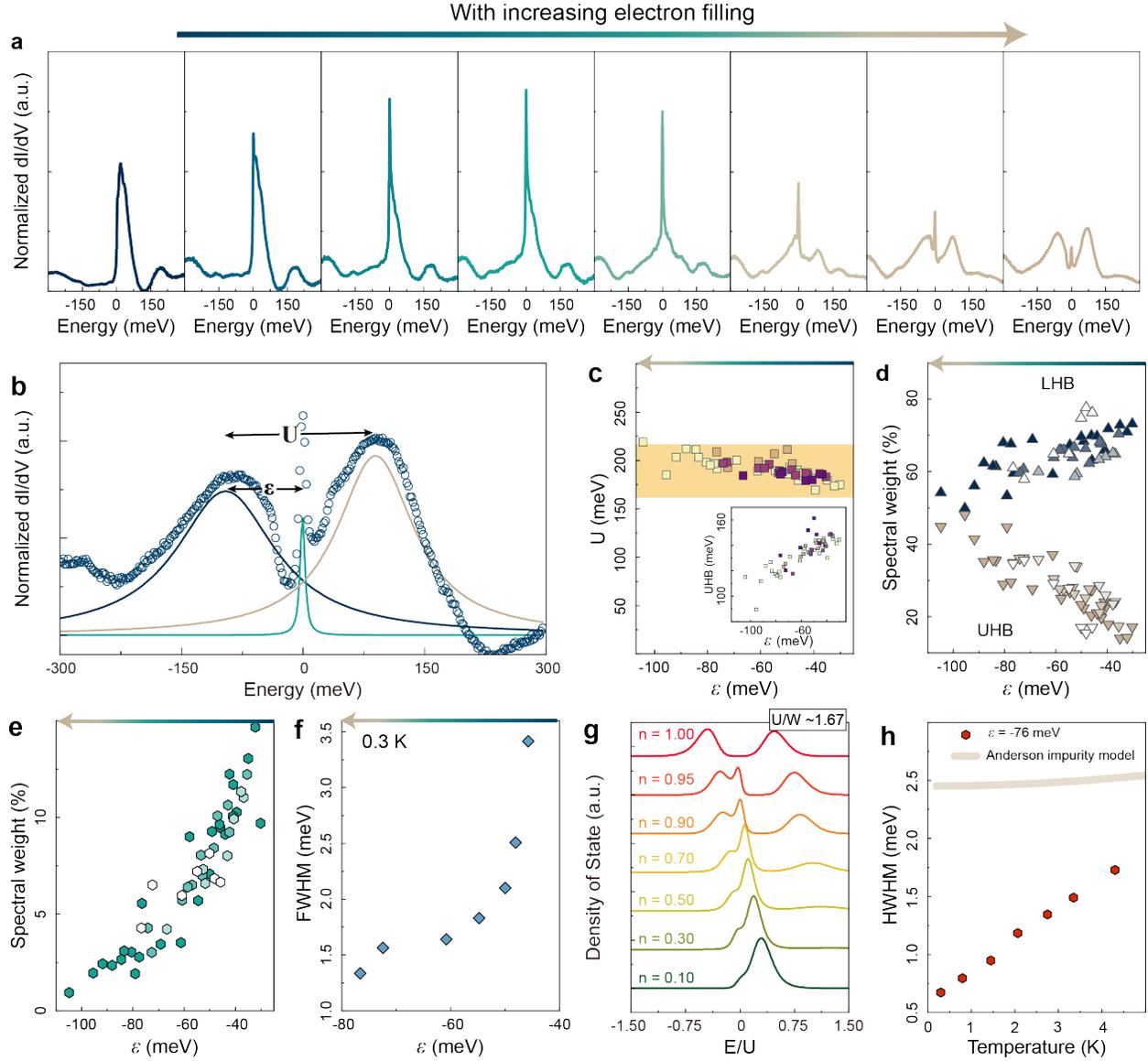

**Fig. 3 Characteristics of the spectral function of a filling-controlled Mott transition. a**, A series of dI/dV spectra selected from Fig. 2d that are sorted by the increased electron filling per cluster. **b**, Typical dI/dV spectrum close to half filling where three peak-like features can be seen. The solid lines are Lorentzian fits to the peaks: dark blue line, LHB; brown line, UHB; cyan line, central narrow quasiparticle peak. **c**, $U$ extracted from the positions of the LHB and UHB, plotted as a function of electron filling (according to the position of the LHB, see SI Fig. S6). The inset shows the positions of LHB and UHB. **d**, Evolution of the spectral weight for LHB and UHB with increasing electron filling. **e**, The spectral weight of the central quasiparticle state with increasing electron filling, and **f**, the FWHM of the central peak. The FWHM is obtained via Fano fitting. **g**, DMFT calculations with quantum Monte-Carlo solvers under the condition $U/W \sim 1.67$. **h**, HWHM of the central quasiparticle peak at different temperatures (see SI Fig. S9). The light gray line denotes the expected temperature-dependent HWHM for a Kondo peak based on the Anderson impurity model ($T_K \sim 40K$).

# Figure 4

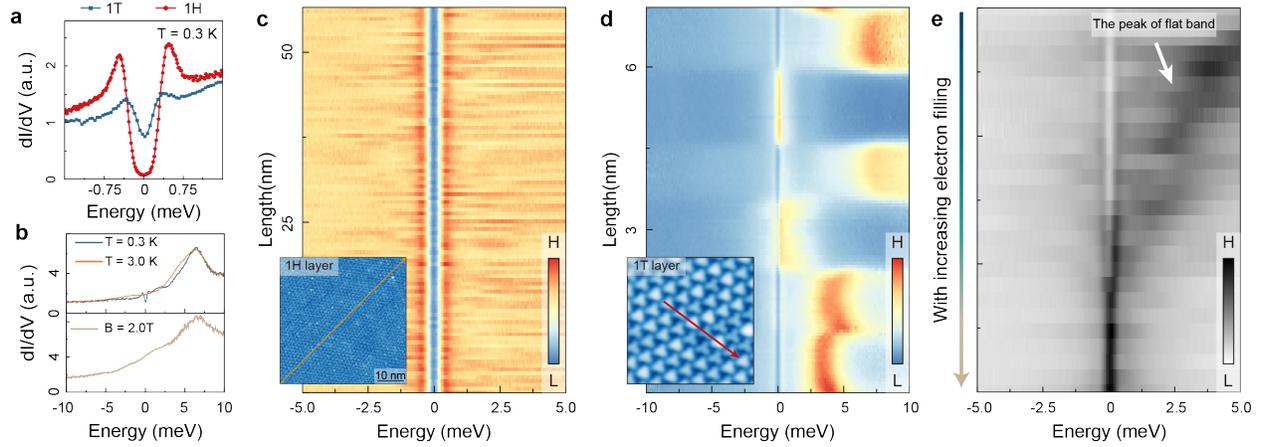

**Fig. 4 The superconductivity dressed by the renormalization of the 1T layer. a**, Superconducting spectra for 1T and 1H terminations at 0.3 K. **b**, Superconducting spectra for 1T termination under different conditions, up: at 3.0 K near the $T_c$, down: at 2 T above the $H_{c2}$. **c, d**, Colormap maps of the spatially resolved dI/dV spectra for 1H termination and 1T termination, with traces marked by orange (**c**) and red (**d**) arrows in their respective inset topography images. **e**, False colormap of superconducting spectra obtained from nearly 30 clusters at the 1T termination and sorting by electron filling.

# Method

**Single crystal growth and characterization:** High-quality single crystals of 4Hb-TaS$_2$ were prepared via the chemical vapour transport method. Appropriate amounts of Ta and S were ground and mixed with a small amount of Se (1% of the S amount). The powder was sealed in a quartz ampoule, and a small amount of iodine was added as a transport agent. The ampoule was placed in a two-zone furnace such that the powder was in the hot zone. After 30 days, single crystals with a typical size of 2.0 mm×1.0 mm×0.1 mm grew in the cold zone of the furnace. The chemical composition of the crystals was determined via energy dispersive X-ray spectroscopy and STM topographic images. Electrical transport measurements were carried out in a Quantum Design physical property measurement system (PPMS-14T). Electrical transport measurement in a conventional 4-lead configuration was realized by attaching four platinum wires to the (001) surface of the sample. The contacts were made with DuPont 4929 N conductor paste.

**STM measurements**. The STM data were acquired with a commercial CreaTec low-temperature system, and the sub-kelvin to 4 K studies were conducted on a Unisoku USM1300 system. For the CreaTec low-temperature STM (Unisoku USM1300), single crystals were cleaved *in situ* under a cryogenic ultrahigh vacuum at T ~ 30 K (T ~ 90 K) and immediately inserted into the STM head. Both Pt/Ir and tungsten tips were used in the experiments. The tungsten tips were annealed to a bright orange color in UHV, and their quality was checked on the surface of single-crystalline Au (111) prior to measurement. Spectroscopic data above 4 K were acquired using the standard lock-in technique at a frequency of 987.5 Hz under a modulation voltage of 2 meV, whereas data below 4 K were obtained using the same technique at a frequency of 987.5 Hz under a modulation voltage of 0.05 ~ 0.2 meV.

**Band calculations and DMFT simulations:** We employ the Vienna ab initio simulation package (VASP) code with the projector augmented wave (PAW) method to perform density functional theory (DFT) calculations. The Perdew-Burke-Ernzerhof (PBE) exchange-correlation functional was used in our calculations. The kinetic energy cutoff is set to 500 eV for expanding the wave functions into a plane-wave basis, and the energy convergence criterion is $10^{-8}$ eV. A Γ-centered $18 \times 18 \times 2$ k-mesh is used in the DFT calculations. After obtaining the band structure from density functional theory calculations, we take the band that crosses the Fermi surface as the non-interacting part of a single-band Hubbard model. In this model, the onsite interaction strength is set to 1.67 W (where W represents the bandwidth). The open-source TRIQS package is employed for dynamical mean field theory (DMFT) calculations, and the continuous-time quantum Monte Carlo algorithm based on hybridization-expansion (CT-HYB) is used to solve the impurity problem. To obtain the spectral function, we adopt the maximum entropy method to analytically continue the Green's function from the Matsubara frequency axis to the real frequency axis. Additionally, the numerical renormalization group (NRG) method is used as a solver for the quantum impurity model, and calculations are performed under the same model parameters as those used in CT-HYB.

## Acknowledgements

We thank Mark H. Fischer, Shichao Yan, Ziji Xiang, Yilin Wang, Haoyu Hu, Junfeng He, Jianjun Ying and Zheng Liu for valuable discussions. This work is supported by the the National Natural Science Foundation of China (Grants No. 12488201 and No. 52261135638), National Key R&D Program of the MOST of China (Grant No. 2022YFA1602600), the Innovation Program for Quantum Science and Technology (Grant No. 2021ZD0302802), the Anhui Initiative in Quantum Information Technologies (Grant No. AHY160000), and the Systematic Fundamental Research Program Leveraging Major Scientific and Technological Infrastructure, Chinese Academy of Sciences under Contract No. JZHKYPT-2021-08.

## Author contributions

Zhenyu Wang and X.H.C. conceived the experiments and supervised this project. P. W., Zhuying Wang and Y. Z. performed STM experiments and data analysis with the assistance of S. Y., W. M., Zeyu L., X. W., J. W., W. C., Z. L., X. Z. and Y. O. M. S., Y. Z., and X. W. synthesized and characterized the samples. Z. C. and K. J. performed theoretical calculations. Zhenyu Wang, K. J., T.W., Y. O. and X.H.C. interpreted the results. Zhenyu Wang, P. W. and X. H. C. wrote the manuscript with input from all authors.

## Competing financial interests

The authors declare no competing interests.

## Data availability

The data supporting the findings of this study are available from the corresponding author upon reasonable request. Source data will be provided with this paper.

## Code availability

The code used for STM data analysis is available from the corresponding author upon reasonable request.